\documentclass[aps,prb,twocolumn,groupedaddress,floatfix,amsmath,showpacs]{revtex4-1}  

\usepackage{graphicx}% Include figure files
\usepackage{url}

\usepackage[colorlinks=true,%
urlcolor=blue,% Webpages
linkcolor=blue,% interne Links
citecolor=blue,% Literaturvz.
hypertexnames=true,% richtige Links
]{hyperref}% Hyperlinking, but without borders
\begin{document}

\title{Spin-Orbit Coupling Effects on Spin Dependent Inelastic Electronic Lifetimes in Ferromagnets}

\author{Steffen Kaltenborn}
\author{Hans Christian Schneider}
\email{hcsch@physik.uni-kl.de}
\affiliation{Physics Department and Research Center OPTIMAS, University of Kaiserslautern, 67663 Kaiserslautern, Germany}


\date{\today}

\begin{abstract}
For the 3d ferromagnets iron, cobalt and nickel we compute the spin-dependent inelastic electronic lifetimes due to carrier-carrier Coulomb interaction including spin-orbit coupling. We find that the spin-dependent density-of-states at the Fermi energy does not, in general, determine the spin dependence of the lifetimes because of the effective spin-flip transitions allowed by the spin mixing. The majority and minority electron lifetimes computed including spin-orbit coupling for these three 3-d ferromagnets do not differ by more than a factor of 2, and agree with experimental results.
\end{abstract}

\maketitle

\section{Introduction}
The theoretical and experimental characterization of spin dynamics in ferromagnetic materials due to the interaction with short optical pulses  has become an important part of research in magnetism.~\cite{Koopmans-NatMat,Muenzenberg-NatMat,Klaeui-NatComm,Rudolf-NatComm,Huebner-NatPhys,Eschenlohr-NatMat} In this connection, spin-dependent hot-electron transport processes in metallic heterostructures have received enormous interest  in the past few years.~\cite{Melnikov-PRL11} In particular, superdiffusive-transport theory has played an increasingly important role in the quantitative interpretation of experimental results.~\cite{Rudolf-NatComm,Eschenlohr-NatMat,Kampfrath-NatNano} Superdiffusive transport-theory, which was introduced and comprehensively described in Refs.~\onlinecite{Battiato-PRLsuperdiff,Battiato-PRBsuperdiff}, uses spin- and energy-dependent electron lifetimes as input,~\cite{Battiato-PRBsuperdiff} and its quantitative results for hot-electron transport on ultrashort timescales in ferromagnetic materials rely heavily, to the best of our knowledge, on the relation between majority and minority electrons for these materials. 

The spin-dependent lifetimes that are used for hot-electron transport, both in ferromagnets and normal metals, are the so-called ``inelastic lifetimes.'' These state (or energy) dependent lifetimes result from out-scattering processes due to the Coulomb interaction between an excited electron and the inhomogeneous electron gas in the system. These lifetimes can be measured by tracking optically excited electrons using spin- and time-resolved 2-photon photoemission (2PPE)~\cite{AeschlimannCo,WeineltCo} and can be calculated as the broadening of the electronic spectral function using many-body Green function techniques.~\cite{Vignale,Onida-RMP02} The problem of the accurate determination of these lifetimes has fueled method development on the experimental and theoretical side,~\cite{Zhukov-PRL93} but has always suffered from the presence of interactions (electron-phonon, surface effects) that cannot be clearly identified in experiment and are difficult to include in calculations. Qualitative agreement was reached for the spin-integrated lifetimes in simple metals and iron,~\cite{Zhukov-PRB06} but even advanced quasiparticle calculations including many-body $T$-matrix contributions, have yielded a ratio between majority and minority lifetimes, which is in qualitative disagreement with experiment for some ferromagnets. A particularly important material in recent studies has been nickel,~\cite{Battiato-PRBsuperdiff,Rudolf-NatComm,Eschenlohr-NatMat} for which the theoretical ratio comes out between 6 and 8,~\cite{Zhukov-PRB06} while the experimental result~\cite{Aeschlimann-FM} is 2. Recent experimental results point toward a similar disagreement for cobalt.~\cite{WeineltCo}

In light normal metals and ferromagnets spin-orbit coupling generally leads to very small corrections to the single-particle energies, i.e., the \emph{band structure}, but it changes the single-particle states qualitatively by introducing a state-dependent spin mixing. With spin-orbit coupling, the average spin of an electron can be changed in transitions due to any spin-diagonal interaction, in particular by electron-phonon momentum scattering.~\cite{Fabian-PRL98,Faehnle-Rapid09,Carva-PRLphonon,Sven-PRB} This is also true for the two-particle Coulomb interaction,~\cite{Krauss-PRB09,Mueller-PRL} as long as one monitors only the average spin of one of the scattering particles, as is done in lifetime measurements by 2-photon photoemission experiments. While this spin mixing due to spin-orbit coupling has recently been included in lifetime calculations for lead,~\cite{Chulkov-PRB11} it was not included in DFT codes used for existing lifetime calculations for 3d-ferromagnets and aluminum,~\cite{Zhukov,Ladstadter,Zhukov-PRB06} whose results are nowadays widely used.

This paper presents results for electron lifetimes in metals and spin-dependent lifetimes in ferromagnets \emph{including spin-orbit coupling}. We show that spin-orbit coupling can be important for electron lifetimes in metals in general. Moreover, the ratio between the calculated majority and minority lifetimes is, for the first time, in agreement with experiment.~\cite{AeschlimannCo,WeineltCo,Aeschlimann-FM} We believe that our calculated electronic lifetimes should be used as an accurate input for calculations of spin-dependent hot-electron dynamics in ferromagnets.

\section{Spin-dependent electron and hole lifetimes in Co and Ni}
We first discuss briefly our theoretical approach to calculate the lifetimes. We start from the dynamical and wave-vector dependent dielectric function $\varepsilon(\vec{q},\omega)$ in the random phase approximation (RPA).~\cite{Vignale,Onida-RMP02,Ladstadter,Zhukov} Our approach, cf. Ref.~\onlinecite{Kaltenborn-PRB13}, evaluates the wave-vector summations in $\varepsilon(\vec{q},\omega)$ without introducing an additional broadening of the energy-conserving $\delta$ function. This procedure removes a parameter whose influence on the calculation for small $q$ is not easy to control and which would otherwise need to be separately tested over the whole energy range. 

\begin{figure}
\includegraphics[width=0.50\textwidth]{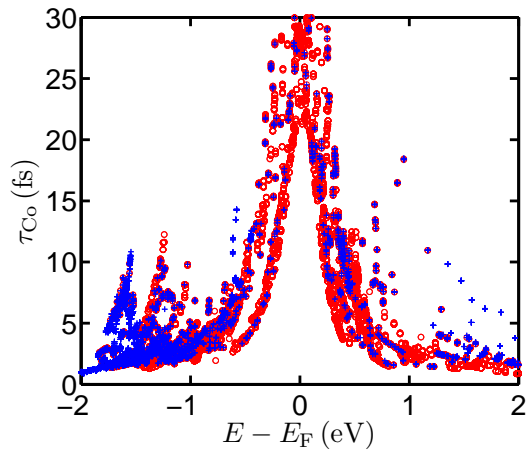}
\caption{\label{fig:lifetimeCo}(Color online) Energy-resolved majority ($\tau^{\uparrow}$, blue $+$) and minority ($\tau^{\downarrow}$, red $\circ$) carrier lifetimes for cobalt. There are in general several different lifetime points at the same energy (see text). We used $17^3$ $\vec{k}$-points in the full BZ.} 
\end{figure}

\begin{figure}
\includegraphics[width=0.50\textwidth]{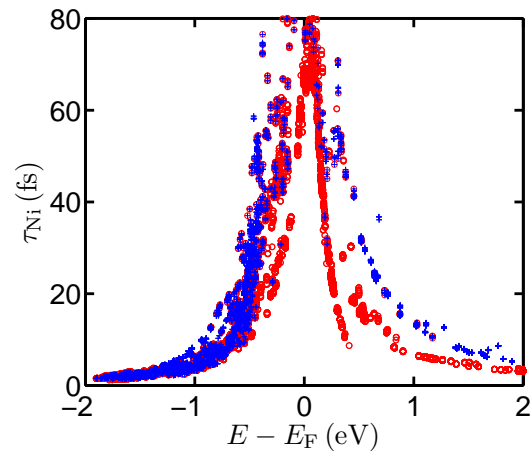}
\caption{\label{fig:lifetimeNi}(Color online) Same as Fig.~\ref{fig:lifetimeCo} for nickel.} 
\end{figure}

The $\vec{k}$- and band-resolved electronic scattering rates, i.e., the inverse lifetimes, $\gamma_{\vec{k}}^{\nu}=(\tau_{\vec{k}}^{\nu})^{-1}$, are calculated using the expression~\cite{Ladstadter,Zhukov} 
\begin{equation}
\gamma_{\vec{k}}^{\nu}=\frac{2}{\hbar}\sum_{\mu\vec{q}}\frac{\Delta q^{3}}{(2\pi)^{3}}V_{q}\big|B_{\vec{k}\vec{q}}^{\mu\nu}\big|^{2}f^{\mu}_{\vec{k}+\vec{q}}\frac{\Im\varepsilon(\vec{q},\Delta E)}{|\varepsilon(\vec{q},\Delta E)|^{2}}.
\label{eq:lifetime}
\end{equation}
Here, the band indices  are denoted by $\mu$ and $\nu$, and $\vec{k}$ and $\vec{q}$ denote wave-vectors in the first Brillouin zone (1.~BZ). The energies $\epsilon_{\vec{k}}^{\mu}$, occupation numbers $f_{\vec{k}}^{\mu}$ and overlap matrix elements $B^{\mu\nu}_{\vec{k}\vec{q}}=\langle \psi_{\vec{k}+\vec{q}}^{\mu}|e^{i\vec{q}\cdot\vec{r}}|\psi_{\vec{k}}^{\nu}\rangle$ are extracted from the ELK DFT (density functional theory) code,~\cite{ELK} which employs a full-potential linearized augmented plane wave (FP-LAPW) basis. Last, $V_{q}=e^2/(\varepsilon_{0}q^2)$ denotes the Fourier transformed Coulomb potential and $\Delta E=\epsilon_{\vec{k}+\vec{q}}^{\mu}-\epsilon_{\vec{k}}^{\nu}$ is the energy difference between initial and final state. For negative $\Delta E$, the distribution function has to be replaced by $-(1-f^{\mu}_{\vec{k}+\vec{q}})$. By using the overlap matrix elements as defined above we neglect corrections due to local field effects. In the language of many-body Green functions, this corresponds to an on-shell $G_{0}W_{0}$ calculation,~\cite{Zhukov,Zhukov-PRB06} where the screened Coulomb interaction ($W_0$) is obtained from the full RPA dielectric function. The $\vec{k}$- and band-dependent wave-functions that result from the DFT calculations including spin-orbit coupling are of the form $|\psi_{\vec{k}}^{\mu}\rangle =a_{\vec{k}}^{\mu}\left|\uparrow\right\rangle +b_{\vec{k}}^{\mu}\left|\downarrow\right\rangle$,~\cite{Fabian-PRL98} where $|\sigma \rangle$ are spinors identified by the spin projection $\sigma=\,\uparrow$, $\downarrow$ along the magnetization direction. According to whether $|a_{\vec{k}}^{\mu}|^2$ or  $|b_{\vec{k}}^{\mu}|^2$ is larger, we relabel each eigenstate by its dominant spin contribution $\sigma$, so that we obtain spin-dependent lifetimes, $\tau_{\vec{k}}^{\sigma}$. Our choice of quantization axis is such that $\sigma=\,\uparrow$ denotes majority carriers states and $\sigma=\,\downarrow$ minority carrier states. Due to the existence of several bands (partially with different symmetries) in the energy range of interest and the anisotropy of the DFT bands $\epsilon^{(\nu)}(\vec{k})$, several lifetimes $\tau_{\vec{k}}^{\nu}$ can be associated with the same spin and energy. When we plot these spin and energy dependent lifetimes $\tau^{\sigma}(E)$ in the following, in particular Figs.~\ref{fig:lifetimeCo} and \ref{fig:lifetimeNi}, this leads to a scatter of $\tau^{\sigma}(E)$ values.

Figures~\ref{fig:lifetimeCo} and \ref{fig:lifetimeNi} display the calculated energy- and spin-resolved carrier lifetimes $\tau^{\sigma}(E)$ around the Fermi energy for cobalt and nickel. The spread of lifetimes at the same energy, which was mentioned above, can serve as an indication for the possible range of results for measurements of energy resolved lifetimes. These ``raw data'' are important for the interpretation of the theoretical results because they already show two important points. First, we checked that there is no good Fermi-liquid type fit to these lifetimes. Second, even if one fits the lifetimes in a restricted energy range by a smooth $\tau(E)$ curve, this ignores the spread of lifetimes, which can be quite sizable as shown in Figs.~\ref{fig:lifetimeCo} and \ref{fig:lifetimeNi}. We believe that such a spread of electronic lifetimes, in particular in the range around 1\,eV above the Fermi energy should be important for the interpretation of photoemission experiments in this energy range, and when these results are used as input in hot-electron transport calculations.

Figure~\ref{fig:lifetimeCo} shows the energy- and spin-resolved lifetimes in cobalt. In addition to the longer lifetimes close to the Fermi energy, hole lifetimes in excess of 5\,fs occur at the top of some d-bands around  $-1.5$, $-1.2$, and $-1\,\mathrm{eV}$. For electronic states with energies above 0.5\,eV longer lifetimes occur at some $\vec{k}$-points. There are also $k$ states with a pronounced spin-asymmetry in the lifetimes (see discussion below). Another important property of cobalt is the existence of two different conduction bands, which intersect the Fermi surface with different slope. This leads to \emph{two} rather well-defined lifetime curves, both for electrons and holes. This can be best seen between $-0.6$ and 0\,eV, where the two curves are shifted by about 0.2\,eV.

The calculated lifetimes in nickel, see Fig.~\ref{fig:lifetimeNi}, do not show a pronounced influence of d-bands and/or anisotropy below the Fermi energy as in cobalt, which is due to the smaller number of bands in the vicinity of the Fermi energy. However, there is a clear spin-dependence of electronic lifetimes, which is most pronounced around 0.4\,eV, but persists almost up to 2\,eV. 

\section{Spin asymmetry of electron lifetimes in Fe, Co, and Ni}
In the following, we will mainly be concerned with lifetimes above 0.3\,eV above the Fermi energy, which is the interesting energy range for the interpretation of photoemission experiments and hot-electron transport calculations, because close to the Fermi energy the influence of phonons is expected to become more pronounced and lead to significantly shorter lifetimes than those predicted by a calculation that includes only the Coulomb interaction. To facilitate comparison with experiment we average the lifetimes in each spin channel in bins of $100\,\mathrm{meV}$ and denote the result by $\bar{\tau}(E)$. The standard deviation of the averaging process then yields ``error bars'' on the $\bar{\tau}(E)$ values. Note that this procedure does not correspond to a ``random $k$'' approximation.

\begin{figure}
\includegraphics[width=0.45\textwidth]{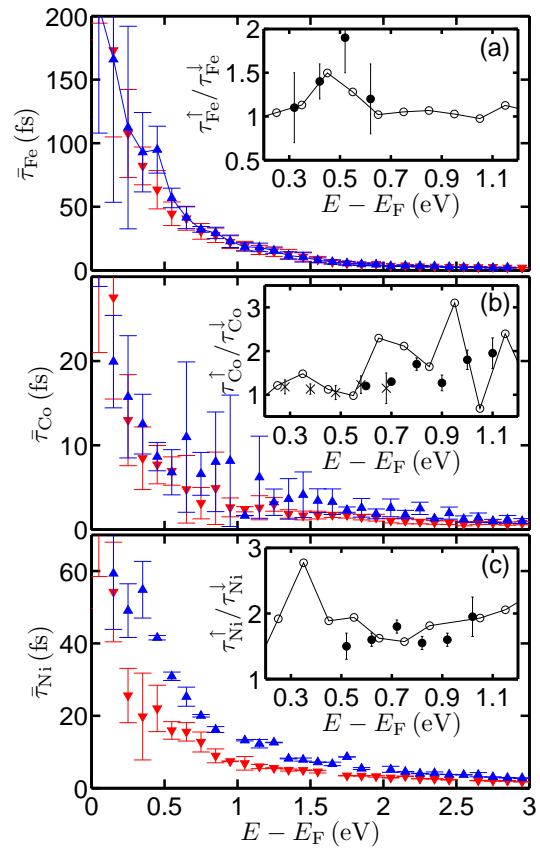}
\caption{\label{fig:lifetimezoom}(Color online) Energetically averaged majority (blue up triangles) and minority (red down triangles) lifetimes for (a) Fe, (b) Co and (c) Ni. The error bars denote the standard deviation obtained from the scatter of the lifetimes as shown in Figs.~\ref{fig:lifetimeCo} and \ref{fig:lifetimeNi}. The insets show the calculated ratio of majority and minority electrons (``$\circ$''), $\tau^{\uparrow}/\tau^{\downarrow}$, in comparison to experimental data, where the ``$\bullet$'' (``$\times$'') correspond to values extracted from Ref.~\onlinecite{AeschlimannCo} and \onlinecite{Aeschlimann-FM} (\onlinecite{WeineltCo}).} 
\end{figure}

Figure~\ref{fig:lifetimezoom} displays the averaged electron lifetimes determined from the data shown in Figs.~\ref{fig:lifetimeCo} and \ref{fig:lifetimeNi}. As insets we have included the ratio of majority and minority lifetimes, $\tau^{\uparrow}/\tau^{\downarrow}$, together with experimental data~\cite{Aeschlimann-FM,AeschlimannCo,WeineltCo} for iron, cobalt and nickel.  Figure~\ref{fig:lifetimezoom}(a) shows that there is only a very weak spin dependence for iron, and the agreement of the ratio~$\tau^{\uparrow} /\tau^{\downarrow}$ with experiment~\cite{Aeschlimann-FM} and recent investigations~\cite{Chulkov,Zhukov-PRL93,Zhukov-PRB06} is quite good, but there is a slight disagreement with earlier, semiempirical studies.~\cite{Penn1,Penn2} However, even an increase of the ratio around $0.5\,\mathrm{eV}$ in the experiment~\cite{Aeschlimann-FM} is well reproduced in our results.

The averaged lifetimes of cobalt, which are shown in Fig.~\ref{fig:lifetimezoom}(b), agree quite well with the experimental lifetimes,~\cite{AeschlimannCo,WeineltCo} but the large error bars extend to a much wider energy range than in iron. This can be traced back to the scatter of lifetimes in Fig.~\ref{fig:lifetimeCo}. The corresponding figure for iron (not shown) exhibits a much smaller scatter. The ratio of majority and minority electron lifetimes, see inset in Fig.~\ref{fig:lifetimezoom}(b),  is around 1 below 0.5\,eV and increases to $\tau_{\uparrow}/\tau_{\downarrow}\simeq2$ for larger energies, a trend that agrees extremely well with measurements.~\cite{Aeschlimann-FM,AeschlimannCo,WeineltCo} To put this result into perspective we note that the experimental data in Ref.~\onlinecite{Aeschlimann-FM} were compared with a theoretical model based on the random $k$ approximation.~\cite{Spicer} If the random-$k$ interaction matrix elements are taken to be spin and energy independent, the majority and minority relaxation times 
are determined by double convolutions over the spin-dependent density-of-states (DOS).~\cite{Zhukov-PRB06} It was found that the experimental results were not in agreement with the ratio of the DOS at the Fermi energy, which led the authors of Ref.~\onlinecite{Aeschlimann-FM} to speculate that the matrix elements for parallel and antiparallel spins should be different due to the Pauli exclusion principle. In our calculations, the effective spin-dependence of the matrix elements is caused exclusively by the spin-mixing due to spin-orbit coupling, but the effect is the same: It makes the ratio of the lifetimes different from the spin-dependent DOS at the Fermi energy.

In Fig.~\ref{fig:lifetimezoom}(c) we turn to nickel. Here,  as in the case of iron, the average lifetimes are slightly larger than the measured ones~\cite{Aeschlimann-FM} (not shown), but due to the small anisotropy in the band structure, the lifetimes in nickel show the smallest error bars and thus an extremely well-defined spin dependence. Only our calculated \emph{majority} electron lifetimes are similar to earlier ab-initio evaluations,~\cite{Chulkov,Zhukov-PRL93,Zhukov-PRB06} but there is an important discrepancy in the ratio  $\tau_{\uparrow}/\tau_{\downarrow}$: The inset of Fig.~\ref{fig:lifetimezoom}(c) shows a ratio of about $\tau_{\uparrow}/\tau_{\downarrow}\simeq 2$, which is independent of energy above 0.4\,eV. This results compares extremely well with experiment, and should be contrasted with the calculated result of Ref.~\onlinecite{Zhukov-PRB06} for $\tau_{\uparrow}/\tau_{\downarrow}\simeq 8$ around 0.5\,eV. These GW calculations (even with a $T$-matrix approach)  gave very similar results to those of the random $k$ approximation~\cite{Zhukov-PRB06} in the energy range above 0.5\,eV. This indicates that the resulting spin asymmetry  $\tau_{\uparrow}/\tau_{\downarrow}\simeq 6$--8 is solely determined by the spin-dependent DOS~$\mathcal{D}_{\sigma}(E)$. Indeed, one has $\mathcal{D}_{\uparrow}(E_\mathrm{F})/\mathcal{D}_{\downarrow}(E_\mathrm{F})\simeq 8$. With the inclusion of spin-orbit coupling, which gives rise to effective spin-flip transitions, the spin asymmetry is no longer determined by the spin-dependent DOS alone. This interpretation is again supported by Ref.~\onlinecite{Aeschlimann-FM} where a strongly enhanced spin-flip matrix element had to be introduced by hand to improve the agreement between a random-$k$ calculation and experiment.

\begin{figure}
\includegraphics[width=0.45\textwidth]{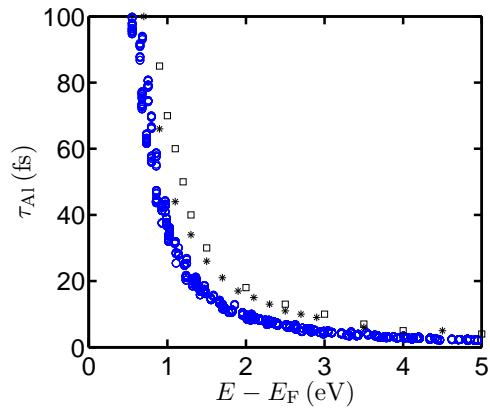}
\caption{\label{fig:lifetimeAl}(Color online) Calculated energy-resolved electronic lifetimes for aluminum (blue ``$\circ$'') in comparison to earlier investigations without spin-orbit interaction. The black squares (stars) correspond to some data extracted from Ref.~\onlinecite{Ladstadter} (\onlinecite{Zhukov-PRB05}). There are in general several different lifetime points at the same energy (see text). We used $17^3$ $\vec{k}$-points in the full BZ.} 
\end{figure}

To conclude the discussion of the ferromagnets, we comment on the spin-integrated lifetimes which can be obtained from the spin-dependent lifetimes, but are not shown here. Compared with experimental lifetimes of Ref.~\onlinecite{Aeschlimann-FM} we generally find an agreement for energies above 0.5\,eV that is on par with earlier calculations.~\cite{Aeschlimann-FM,Chulkov,Zhukov-PRL93,Zhukov-PRB06} For energies below 0.5\,eV where the error bars on the averaged lifetimes are largest, the calculated lifetimes are larger than the measured ones, but in this energy range a good agreement with experiments cannot be expected because of scattering processes, which appear as elastic due to the energy resolution of the photoemission experiments.

\section{Influence of spin-orbit coupling on electron lifetimes in Al}
To underscore the importance of spin-orbit coupling for lifetime calculations, we also briefly discuss our calculated results for electronic lifetimes in aluminum in comparison with earlier investigations~\cite{Ladstadter,Zhukov-PRB05} without spin-orbit effects. Fig.~\ref{fig:lifetimeAl} shows that  smaller electronic lifetimes result for aluminum when spin-orbit coupling is included. In particular in the energy range between 1 and 3\,eV the lifetimes differ by almost a factor of two. Thus the inclusion of spin-orbit coupling improves the agreement with experiment (see, for instance, Ref.~\onlinecite{Bauer}), which was already quite good for the existing calculations.~\cite{Zhukov-PRB05} It is conceivable that the use of more sophisticated many-body techniques, such as the inclusion of vertex corrections or using a $T$-matrix approach,~\cite{Zhukov-PRB05} might lead to further improvements. As in the case of the ferromagnets, the electronic band structure is practically unchanged by the inclusion of spin-orbit coupling, but the rather large effect of the spin-orbit coupling on spin relaxation in aluminum through spin hot-spots has already been demonstrated.~\cite{Fabian-PRL98} Another argument for the importance of the spin-orbit coupling is that the spin-mixing allows transition between the Kramers degenerate bands. These transitions between Kramers degenerate bands may have a remarkable influence even on electron-gas properties that are usually assumed to be spin-independent, such as the intraband plasma frequency.~\cite{Kaltenborn-PRB13}

\section{Conclusion}
In conclusion, we presented ab-initio results for spin-dependent electronic lifetimes in ferromagnets and aluminum including spin-orbit coupling. We found that the electronic lifetimes in iron exhibit no visible spin dependence in the range of $-2$ up to $3\,\mathrm{eV}$ in agreement with earlier results, whereas the ratio $\tau^{\uparrow}/\tau^{\downarrow}$ between majority and minority lifetimes does not exceed 2 for cobalt and nickel. Our results agree well with experimental data, but differ from earlier calculations, which found that $\tau^{\uparrow}/\tau^{\downarrow}$ was essentially determined by the spin-dependent density-of-states. We showed that, by allowing for effectively spin-changing transitions as contributions to the lifetime, spin-orbit coupling is the essential ingredient that can make the spin asymmetry of the electronic lifetimes much smaller than the spin-asymmetry of the density-of-states. Inclusion of our calculated spin dependent lifetimes in transport calculations should make it possible to more accurately characterize the influence of spin-dependent hot-electron transport on magnetization dynamics.

\begin{acknowledgments}
We are grateful for a CPU-time grant from the J\"{u}lich Supercomputer Centre (JSC). We acknowledge helpful discussions with M. Aeschlimann, M. Cinchetti, and S. Mathias.
\end{acknowledgments}

\end{document}